# Sampling Time Effects for Persistence and Survival in Step Structural Fluctuations

D.B. Dougherty[1], C. Tao[1], O. Bondarchuk[18], W.G. Cullen[1], and E.D. Williams[1,2]
Department of Physics[1] and
Institute for Physical Science and Technology[2]
University of Maryland
College Park, MD 20742-4111

and

M. Constantin, C. Dasgupta**, S.Das Sarma
Condensed Matter Theory Center
University of Maryland
College Park MD, 20742-4111

The effects of sampling rate and total measurement time have been determined for single-point measurements of step fluctuations within the context of first-passage properties. Time dependent STM has been used to evaluate step fluctuations on Ag(111) films grown on mica as a function of temperature (300-410 K), on screw dislocations on the facets of Pb crystallites at 320K, and on Al-terminated Si(111) over the temperature range 770K - 970K. Although the fundamental time constant for step fluctuations on Ag and Al/Si varies by orders of magnitude over the temperature ranges of measurement, no dependence of the persistence amplitude on temperature is observed. Instead, the persistence probability is found to scale directly with $t/\Delta t$ where $\Delta t$ is the time interval used for sampling. Survival probabilities show a more complex scaling dependence, which includes both the sampling interval and the total measurement time $t_m$. Scaling with $t/\Delta t$ occurs only when $\Delta t/t_m$ is a constant. We show that this observation is equivalent to theoretical predictions that the survival probability will scale as $\Delta t/L^z$, where $L$ is the effective length of a step. This implies that the survival probability for large systems, when measured with fixed values of $t_m$ or $\Delta t$ should also show little or no temperature dependence.

**Introduction**

Direct imaging of spatial distributions and temporal variations of structures on surfaces has provided a remarkable test bed for experimental application of fundamental statistical mechanics [1, 2]. In addition, non-thermodynamic stochastic properties related to the first passage problem [3] can be evaluated directly from such experimental observations [4-7]. The first passage problem, which involves determining the distribution of times for the first return of a stochastic process to its starting point, is experimentally difficult to measure because of the intrinsically low statistics. The persistence probability, which is essentially an integral of the first passage distribution, measures the probability of a random process NOT returning to its original configuration in a time interval t. A related quantity, the survival probability measures the probability of NOT reaching a fixed reference point in a time interval t. While the two definitions appear similar, the persistence and survival probabilities are profoundly different in their relationship to the physical properties underlying the stochastic process [7]. Correctly interpreting the physical meaning of experimental observations of persistence and survival requires careful attention to the details of experimental design [8-11]. In this work, we explicitly investigate the issues of the sampling interval (which generally corresponds to substantial under sampling with respect to the physical time constants of the system) and the total measurement time. We show that these turn out to have subtle importance in the persistence properties of step fluctuations. As a test system, we have measured step fluctuations (e.g. x(t), the step position vs. time) on Ag(111) thin films and on Al-terminated Si(111) surfaces over a temperature range for which the time scale of the physical fluctuations varies dramatically [2, 8, 12-15]. We have also measured step fluctuations on room-temperature Ag thin films and on screw dislocations on a Pb crystallite, as shown in Fig. 1a, with variations in the time protocols for the measurement designed to test theoretical predictions.

Step fluctuation mechanisms for all of these systems have been determined previously by using the measured step position x(t) to determine the time correlation function and extract from it a characteristic dynamic exponent, z. This function is defined by $G(t) = \left\langle (x(t + t_0) - x(t_0))^2 \right\rangle$, where the brackets indicate an average over all initial times, $t_0$. For short enough times, this function grows as $t^{1/z}$, where z = 2 indicates rate limiting attachment/detachment at the step edges and z = 4 indicates rate limiting diffusion along the step edge [1, 2]. For the Al/Si surface, z = 2 scaling is observed [15, 16], while for both Pb(111) and Ag(111) z = 4 scaling is observed [5, 6, 8]. While various more complicated mechanisms[17, 18] often need to be considered to explain the observed scaling, they are not believed to be relevant for any of the systems described herein [19, 20].

Our previous measurements have shown that for Ag(111), the time constant for step-edge diffusion ranges from $10^{-4}$ s at room temperature to $10^{-7}$ s at 450K [8]. For Al/Si(111), the time constant for attachment/detachment at the step edge ranges from a quarter of a second at 770K to 0.3 ms at 1020K [15]. The strong temperature-induced variation of the underlying physical time constant in these systems allows a wide dynamic range of experimental sampling time relative to the physical time to be evaluated. By additionally varying the temporal measurement protocol at fixed temperature, we can distinguish time sampling effects from physical effects. We have used steps on Pb(111) and Ag(111), both in the class of step-edge diffusion-limited step motion, for these tests.

Our previous results have shown that the time correlation function is independent of sampling time interval Δt (the time between sequential measurements of the step position x), although the effective system size is limited by the overall measurement time $t_m$ for highly ordered systems [8]. However, as will be shown here, measurements of the first-passage properties of fluctuations, e.g. the persistence or survival probabilities for step wandering are strongly dependent on the measurement protocol. We have previously confirmed experimentally [5,6] the prediction that the persistence probability follows a simple power law at early times

$$P(t) \sim t^{-\theta} \tag{1}$$

where the experimentally measured persistence exponent $\Theta$ is, consistent with theoretical predictions, equal to 3/4 or 7/8 for attachment/ detachment (z = 2) or step-edge diffusion (z = 4) mechanisms respectively. Here we will show that the absolute magnitude of the persistence probability depends only on the sampling interval $\Delta t$ of measurement. The survival probability is related to the auto-correlation function [7], but unlike the auto-correlation function it does not scale simply with the system size L. Instead, the survival probability is predicted to scale with either system size or sampling interval, but only when the ratio $\Delta t/L^z$ is a constant. We are able to test this non-trivial prediction systematically due to our prior observation that the effective system size is determined by the total measurement time [8]. By using independently variable total measurement time and sampling interval, we demonstrate that time scaling of the survival probability does indeed occur only for the predicted scaling ratio of sampling interval and system size.

**Experimental**

The sample preparation techniques for Al/Si(111) [15,16], Pb crystallites [21-23] and Ag(111) thin films [8] have been presented previously. For observation of step fluctuation we use repeated STM scans across a step boundary, as shown for a screw dislocation on Pb(111) in Fig.

1b.  The sampling time interval Δt is equal to the time between scans, and the total measurement time $t_m$ is the number of lines times the time interval.  For the temperature dependent Ag(111) measurements, the total measurement times were 26.9 s and 39.4s with 512 line scans.  Ten to twenty such samples were measured for each set of experimental conditions.  For Al/Si(111), earlier measurements were reported for measurement times of 33 to 38 seconds with 512 line scans. The numbers of data sets averaged were 11 at 970K, 6 at 870K, and 3 at 770K.  Here we also show measurements with total measurement times of 105s for T=770K and 870K.  For Pb(111) the total measurement times were 19s and 177 s with 512 line scans.  Five data sets were averaged for the 19s curves and 4 were averaged for the 177s curves.  Room temperature measurements of Ag(111) were also performed with total measurement times of 25s and 100 s and sampling intervals of 0.05s and 0.20s.

The step position x(t) was extracted from each line scan after flattening the overall image.  In some cases, the step edge was identified as the point at which surface height was midway between the heights of the upper and lower terraces.  In other cases, the point of inflection was used to identify the step edge. The individual *x(t)* data sets are used to calculate individual correlation functions, and the reported correlation functions and probability distributions are averaged over the individual measurements.  Error bars reported are the standard deviation (one-sigma) and are obtained from weighted fits or from the deviation of repeated measurements.  The persistence and survival probabilities are calculated by dividing the data into time bins, where the sampling interval Δt sets the smallest time bin , $t_{min}$ = Δt = measurement time/number of line scans.  Each time bin was evaluated as "persistent" if the measured positions in the time bins were all positive (or all negative) with respect to the position at the start of the time bin, and "non-persistent" otherwise.  The tabulation of the fraction of persistent bins for each bin width t yields the persistence probability P(t). For survival, the analysis procedure is similar, except that the reference position is the average (over the entire measurement time $t_m$) of the step position, rather than the step position at the beginning of each time bin.

**Results and Analysis**

The strong dependence of the step fluctuations on temperature for steps on Al/Si(111) is illustrated with plots of measured position vs. time in Fig. 2. Analysis of the persistence probability has previously shown [5, 6] power law behavior with an exponent of $\Theta = 0.77 \pm 0.03$. In Figure 3a, we show explicitly the dependence of the measured persistence on the sampling interval $\Delta t$. The persistence probability is the same for data measured at different temperatures, but with the same $\Delta t$. It is different for data measured at the same temperature, but with different $\Delta t$. The effect of scaling the time axis to $t/\Delta t$ is shown in Fig. 3b.

We have also evaluated the effect of temperature on measured persistence for steps on Ag(111). In Fig. 4, the persistence probabilities measured over a wide range of temperature, with the same measurement protocol ($t_m = 26.9$s, $\Delta t = 0.053$s) are shown, with one data set measured with $t_m = 39.3$s, $\Delta t = 0.076$s. The persistence exponents measured at all temperatures scatter about the expected value of $\Theta = 7/8$, with no systematic temperature dependence. The scatter in the value of the exponent yields some deviation in the curves, but the only significant offset occurs for the T=140°C data measured with a different time interval. Rescaling this data by the ratio of the measurement intervals eliminates the offset, as shown by the solid line in Fig. 4.

These results for the scaling of persistence are consistent with numerical simulations of persistence behavior as shown in Figure 5. We have investigated these aspects in a detailed numerical study in which a simple Euler scheme [4, 24] is used to numerically integrate a spatially discretized version of the fourth order conserved Langevin equation

$$\frac{\partial x(y,t)}{\partial t} = -\frac{\Gamma_h \tilde{\beta}}{kT}\frac{\partial^4 x(y,t)}{\partial y^4} + \eta(y,t) \qquad (2)$$

with $\left\langle \eta(y,t)\eta(y',t')\right\rangle = -2\Gamma_h \nabla_y^2 \delta(y-y')\partial(t-t')$, $\tilde{\beta}$ equal to the step stiffness, and $\Gamma_h$ equal to the step mobility [1,25]. This equation provides an excellent description of step fluctuations governed by diffusion along the edge of the step. The results shown here were obtained in the equilibrium regime. As shown in the figure, for a fixed system size, the persistence probability measured with discrete sampling intervals Δt is shifted along the time axis as Δt changes. Normalizing the measurement time to t/Δt causes all the curves to collapse to a single curve, as also observed experimentally. Similar numerical calculations in which system size is changed while the time interval is fixed yield *no shift* in the persistence measurement.

We have performed a more rigorous experimental test of the effects of time scaling for step fluctuations on Pb(111). In these measurements the total measurement time $t_m$ and the measurement interval Δt are changed by nearly a factor of ten while maintaining a constant value of $t_m$/Δt =512. The time correlation function,

$$G(t) = \left\langle \left(x(t)-x(0)\right)^2\right\rangle = \left(\frac{2\Gamma(1-1/n)}{\pi}\right)\left(\frac{kT}{\tilde{\beta}}\right)^{\frac{n-1}{n}}\left(\Gamma_n t\right)^{\frac{1}{n}}, \qquad (3)$$

is shown in Fig. 6a for measurements with different measurement time ($t_m$ = 19s and 177s). Consistent with previous studies [5,26] clear $t^{1/4}$ time scaling is observed, as expected for step-edge diffusion mediated fluctuations (n=4 in Eq. 3). The short-time behavior of the time correlation is clearly unaffected by the differences in measurement protocol. The persistence behavior, however, is strongly dependent on the measurement interval, with the curves collapsing when scaled as a function of t/Δt.

Time correlation and persistence both have a functional dependence on relative step displacements (e.g. position at time t relative to position at time t=t').  The autocorrelation function and the survival probability, in contrast, depend on the absolute step positions, which are generally referenced to the average step position.  In Fig. 7, we show these functions measured for the two different measurement protocols of Pb(111).  The strong dependence of the autocorrelation on measurement time is illustrated in Fig. 7a.  We can quantify the effects by comparison with the predicted form for the autocorrelation function for step fluctuations:

$$C(t) = C(0)\left[\exp(\frac{-t}{\tau_c}) - (\frac{t}{\tau_c})^{\frac{1}{n}}\Gamma(\frac{n-1}{n},\frac{t}{\tau_c})\right] \qquad (4a)$$

$$C(0) = \frac{k_B T L_{eff}}{2\pi^2\tilde{\beta}} = w_{eq}^2 \qquad (4b)$$

$$\tau_c = \frac{k_B T}{\Gamma_h \tilde{\beta}}\left(\frac{L_{eff}}{2\pi}\right)^z, \qquad (4c)$$

where $L_{eff}$ is the effective system size.  The early time behavior of C(t) (i.e. letting t approach 0) for step--edge diffusion limited fluctuations, n=4, is:

$$C(t) = C(0)\left[1 - (\frac{t}{\tau_c})^{\frac{1}{z}}\Gamma(\frac{3}{4})\right] \qquad (5)$$

Fits of the data are shown in Fig. 7a as solid lines.  The fits yield values of C(0) = 149Å$^2$ and $\tau_c$ = 21.3 s for the measurement time of 177s, and values of C(0) = 75 Å$^2$ and $\tau_c$ = 3.4s for the measurement time of 19s.  The ratio of $t_m/\tau_c$ of 6-8 is consistent with our previous observations on Ag(111).

The results of analyzing the step fluctuations for survival are shown in Figs 7b and 7c. The survival probabilities measured with the different time protocols are distinctly different, with the longer measurement time (and longer measurement interval) yielding a much slower decrease with time, e.g. a larger apparent survival time constant.  The data scale well at short times with

$t/\Delta t$, as shown in Fig. 7c.  As we will discuss below, the constant ratio of measurement time and

time interval used in this measurement is a significant factor in this scaling result.

We evaluated the effects of varying the measurement time and sampling interval independently with measurements on steps on Ag(111) at room temperature.  Measurement protocols of $t_m = 100s$, $\Delta t = 0.05s$, $t_m = 100s$, $\Delta t = 0.20s$, and $t_m = 25s$, $\Delta t = 0.05s$ were used to

allow pairwise comparisons of measurements with the same total time and different sampling

interval, and measurements with different total time and the same sampling interval.  The results

for the autocorrelation function are shown in Fig. 8.  The immediate qualitative evaluation that

the correlation time depends strongly only on the total measurement time is confirmed by

quantitative fits to the functional form of Eq. 4 with n =4.  The results yield:

$t_m = 25s$, $\Delta t = 0.05s$,    $C(0) = 0.015 \pm 0.004$,  $\tau_c = 3.5 \pm 1.6$ s

$t_m = 100s$, $\Delta t = 0.05s$,  $C(0) = 0.021 \pm 0.003$,  $\tau_c = 13.1 \pm 2.2$ s

$t_m = 100s$, $\Delta t = 0.20s$,  $C(0) = 0.028 \pm 0.005$,  $\tau_c = 13.0 \pm 2.5$ s

Within experimental uncertainty, the curves measured with the same total measurement time are identical despite different sampling intervals, while those measured with different measurement times are significantly different even when measured with the same sampling interval.

The dependence of the survival probability on measurement protocol is different, as shown in Fig. 9.  Figure 9a shows the three survival curves (corresponding to the autocorrelation functions of Fig. 8 measured under the same conditions) as a function of scaled time $t/\Delta t$.  The result shows that the two curves for which the ratio $t_m/\Delta t$ is a constant (=500) collapse to a single curve over the same time range where the autocorrelation is well fit by the analytical form of Eq.

4.  Figure 9b shows the same data replotted as a function of $t/t_m$.  The scaling dependence is the

same – data collapse only occurs for the same two curves with the constant ratio of $t_m/\Delta t$.  These results show unambiguously that the survival probability does not scale independently with either the sampling interval or the total measurement time. Instead scaling requires a fixed ratio of these two quantities. This is confirmed by numerical simulations designed to test the same effects of measurement time and sampling interval. As shown in Fig. 10, the same scaling dependence is observed as for the experiments for calculations where a finite measurement time limits calculations of the average properties. These simulations were performed for the spatially discretized one-dimensional Edwards-Wilkinson equation (z=2). Measurements of the survival probability were carried out for equilibrated samples with L=200 sites and periodic boundary conditions. The measured average of the height at each site over the measurement time $t_m$ was used as the reference point in the calculation of the survival probability, and the results were averaged over $10^4$ independent runs.

**Discussion and Conclusions**

The dependence of the persistence probability on discrete sampling time, $\Delta t$, has been studied theoretically by Majumdar *et al.* [9]. In this work, the important result that under sampling should not change the value of the persistence exponent was obtained. Later Ehrhardt *et al.* [10] noted in passing that discrete time sampling should nevertheless change the absolute magnitude of the persistence probability. The experimental results presented above clearly verify these conclusions and can be understood quite simply from the point of view of normalization. For the smallest time interval $\Delta t$, changes in the step position cannot be observed and therefore *return* to an initial configuration is impossible. Thus, $p(\Delta t)$ is constrained to be unity. With different sampling times it is then obvious that the probability curves will be shifted along the time axis by changes in $\Delta t$. As a result the magnitude of $p(t)$ will be apparently different, and differences of the curves will collapse when scaled by t/Δt. As discussed in Constantin *et al.* [27] the persistence probability at long times also depends on the scaling variable $\Delta t/L^z$ where L is the "true" length of the step. However, this dependence is observed only for values of t large compared to the correlation time, which is proportional to $L^z$. Since the measurement times in the experiments and simulations described here are small compared to the true correlation time, the dependence on the scaling variable $\Delta t/L^z$ is not observed, and the data for the persistence probability exhibit simple scaling in t/Δt.

Interestingly, the dependence of the survival probability on Δt is more subtle. In Ref. 7, Dasgupta *et al.* justified numerically the following scaling form for the survival probability

$$S(t,L,\Delta t) \sim f(\frac{t}{L^z},\frac{\Delta t}{L^z}) \qquad (6)$$

where Δt is the sampling interval, t is time, and L is the length of the step. The fundamental observation for *S(t)* presented in this work, namely that it scales with Δt only if $\Delta t/t_m$ is constant, is a simple consequence of Eq. 6. This becomes obvious when the effective length of a step edge as outlined in Ref. 8 is recalled. For very long steps, analysis of experimental correlation times indicates that the effective length of the step is set by the longest wavelength fluctuation that has time to decay during the observation time $t_m$. Put simply,

$$L_{eff} \sim \lambda_{\max} \sim t_m^{1/z} \qquad (7).$$

Thus it is clear that, in the scaling function from Eq. 6, constant $\Delta t/t_m$ corresponds simply to a constant value of the second argument, $\dfrac{\Delta t}{L^z}$. Only then does the survival probability scale simply with Δt.

In summary, we have shown experimental results for a wide variety of systems that illustrate that the magnitude of the persistence probability scales with sampling time even though the persistence exponent is independent of sampling time. This effect completely obscures any temperature dependence of the persistence probability. The scaling of the related survival probability is more complicated and can be traced back to theoretical predictions in Ref. 7. For the same systems, survival only shows simple scaling with sampling time for a constant ratio of sampling time to total measurement time.

The observations presented in this work were initiated by an attempt to understand the dependence of persistence and survival probabilities on material-dependent thermodynamic and kinetic parameters governing surface mass transport. It is clear that such information is often obscured by the experimental necessity of discrete time sampling, and further that incorrect conclusions could be drawn if the effects of discrete sampling are not evaluated properly. The converse problem of understanding the physical significance of persistence and survival probabilities will clearly depend on the meaning of the sampling interval and the measurement time in the chosen application scenario [28-31].

**Acknowledgment**

This work has been supported by the NSF-NIRT under grant DMR-0102950 with partial support from U.S. Department of Energy Award No. DOE-FG02-01ER45939 and from the US-ONR. We also gratefully acknowledge support and SEF support from the NSF MRSEC under grant DMR 00-80008.

## References

1    H.-C. Jeong and E. D. Williams, Surface Science Reports **34**, 171 (1999).

2    M. Giesen, Progress in Surface Science **68**, 1 (2001).

3    S. Redner, *A Guide to First-Passage Processes* (Cambridge University Press, Cambridge, 2001).

4    J. Krug, H. Kallabis, S. N. Majumdar, S. J. Cornell, A. J. Bray, and C. Sire, Physical Review **B56**, 2702 (1997).

5    D. B. Dougherty, O. Bondarchuk, M. Degawa, and E. D. Williams, Surface Science **527**, L213 (2003).

6    D. B. Dougherty, I. Lyubinetsky, E. D. Williams, M. Constantin, C. Dasgupta, and S. Das Sarma, Physical Review Letters **89**, 136102 (2002).

7    C. Dasgupta, M. Constantin, S. Das Sarma, and S. N. Majumdar, Physical Review E **69**, 022101 (2004).

8    O. Bondarchuk, D. B. Dougherty, M. Degawa, M. Constantin, C. Dasgupta, and S. Das Sarma, submitted (2004).

9    S. Majumdar, A.J. Bray, and G. C. M. A. Erhardt, Physical Review E **64**, 015101R (2001).

10   G. C. M. A. Erhardt, A.J. Bray, and S. N. Majumdar, Physical Review E **65**, 041102 (2002).

11   R. K. P. Zia and B. Schmittmann, American Journal of Physics **71**, 859 (2003).

12   K. Morgenstern, G. Rosenfeld, B. Poelsema, and G. Comsa, Physical Review Letters **74**, 2058 (1995).

13   K. Morgenstern, G. Rosenfeld, G. Comsa, M. R. Sørensen, B. Hammer, E. Lægsgaard, and F. Besenbacher, Physical Review B **63**, 045412 (2001).

14   J.-M. Wen, S.-L. Chang, J. W. Burnett, J. W. Evans, and P. A. Thiel, Physical Review Letters **73**, 2591 (1994).

15   I. Lyubinetsky, D. B. Dougherty, T. L. Einstein, and E. D. Williams, Physical Review B **66**, 085327 (2002).

16   I. Lyubinetsky, D. Daugherty, H. L. Richards, T. L. Einstein, and E. D. Williams, Surface Science **492**, L671 (2001).

17   T. Ihle, C. Misbah, and O. Pierre-Louis, Physical Review B **58**, 2289 (1998).

18   S. V. Khare and T. L. Einstein, Physical Review **B57**, 4782 (1998).

19   T. L. Einstein, Surface Science **521**, L669 (2002).

20   D. B. Dougherty, I. Lyubinetsky, T. L. Einstein, and E. D. Williams, Physical Review B, in press (2004).

21   K. Arenhold, S. Surnev, H. P. Bonzel, and P. Wynblatt, Surface Science, submitted (1998).

22   A. Emundts, H. P. Bonzel, P. Wynblatt, K. Thürmer, J. Reutt-Robey, and E. D. Williams, Surface Science **481**, 13 (2001).

23   D. B. Dougherty, K. Thürmer, M. Degawa, W. G. Cullen, J. E. Reutt-Robey, and E. D. Williams, Surface Science **554**, 233 (2004).

24   M. Constantin, S. D. Sarma, C. Dasgupta, O. Bondarchuk, D. B. Dougherty, and E. D. Williams, Physical Review Letters **91**, 086103 (2003).

25   N. C. Bartelt, J. L. Goldberg, T. L. Einstein, and E. D. Williams, Surface Science **273**, 252 (1992).

26    L. Kuipers, M. S. Hoogeman, J. W. M. Frenken, and H. van Beijeren, Physical Review B **52**, 11387 (1995).

27    M. Constantin, C. Dasgupta, P. P. Chatraphorn, S. N. Majumdar, and S. Das Sarma, Physical Review E **69**, 061608 (2004).

28    Z. Toroczkai and E. D. Williams, Physics Today **52**, 24 (1999).

29    H. S. Wio and K. Lindenbert, in *AIP Conference Proceedings: Modern Challenges in Statistical Mechanics: Patterns, Noise and the Interplay of Nonlinearity and Complexity*, edited by V. M. Kenkre and K. Lindenberg (AIP, New York, 2003), Vol. 658.

30    R. van Zon, S. Ciliberto, and E. G. D. Cohen, Physical Review Letters **92**, 130601 (2004).

31    E. D. Williams, Materials Research Society Bulletin **September**, 621 (2004).

**Figure captions**

Fig.1: (color online) a.) 475 nm x 475 nm STM image (differentiated for clarity) of a screw dislocation emerging on the top facet of a Pb crystallite supported on Ru(0001) measured at 320 K. Tunneling conditions are: U=0.34V, $I_t$= 0.03 nA, b) Pseudo image obtained by scanning the STM tip repeatedly across a single line perpendicular to the step edge with the same tunneling conditions as in a.). Vertical axis shows 512 line scans initiated at equal sampling intervals over 24 s along a horizontal distance of 67.2nm.

Figure 2: (color online) Measurements of step edge position vs. time for steps on Al/Si(111) with arbitrary offsets for clarity. Data points are spaced by 68 ms for the 970 K and 870 K steps and by 74 ms for the 770 K step.

Fig. 3 (color online) Al/Si(111) measured persistence for three different temperatures and two different time-sampling conditions. a) p(t) is shown vs. t, revealing that measurements with the same sampling interval are indistinguishable b) persistence is plotted as a function of $t/\Delta t$, showing complete collapse of all the data sets.

Figure 4 (color online) Experimentally determined persistence probability *P(t)* vs. *t* for a range of temperature for indirectly heated Ag films, calculated as described in the text (Experimental section). The data shown were measured with $\Delta t = 0.051$ s for all temperatures except T = $140^o$C, where $\Delta t = 0.076$ s. The solid line shows the T=$140^o$C data vs. time (upper axis) scaled by a factor of 0.051/0.076.

Figure 5 –(color online) Calculation demonstrating persistence scaling. Three numerical simulations run with the same system size, but different sampling intervals $t_{sample}$. Scaling as $t/t_{sample}$ is demonstrated.

Figure 6: (color online) Pb(111) at 320K. Blue – total sampling time 19 s, sampling time interval $\Delta t$ =37 ms. Red – total sampling time 177 s, sampling time interval $\Delta t = 0.346$ s. Time

correlation function G, persistence probability p and survival probability, and persistence probability scaled as t/ Δt.

Fig. 7: (color online)  Pb(111) at 320K.   Blue – total sampling time 19 s, sampling time interval Δt =37 ms.    Red – total sampling time 177 s, sampling time interval Δt = 0.346 s. a) Autocorrelation function C, data = circles, solid lines are fits to Eq. 5.  b)  survival probability S and c)  survival probability scaled as t/ Δt.

Figure 8: (color online)  Correlation function, C(t), for Ag(111) at room temperature obtained with different measurement protocols

Figure 9: (color online) Survival probability, S(t), for Ag(111) at room temperature with the same three different measurement protocols as in Fig. 8.  Upper graph, time scaled with sampling interval.  Lower graph, time scaled with total measurement time.

Figure 10: Calculated survival probability for the one dimensional Edwards-Wilkinson equation (z=2), with L = 200 sample for measurement time $t_m$ = 100 using a sampling interval Δt = 1, and for $t_m$ = 400, using Δt = 1 and 4.  The measured average of the height at each site over the measurement time $t_m$ was used as the reference point in the calculation of the survival probability. The survival probability is shown as a function of the scaled time t/Δt.

Solid line:        $t_m$ = 100, Δt = 1

Triangles:        $t_m$ = 400, Δt = 4

Circles:        $t_m$ = 400, Δt =1.

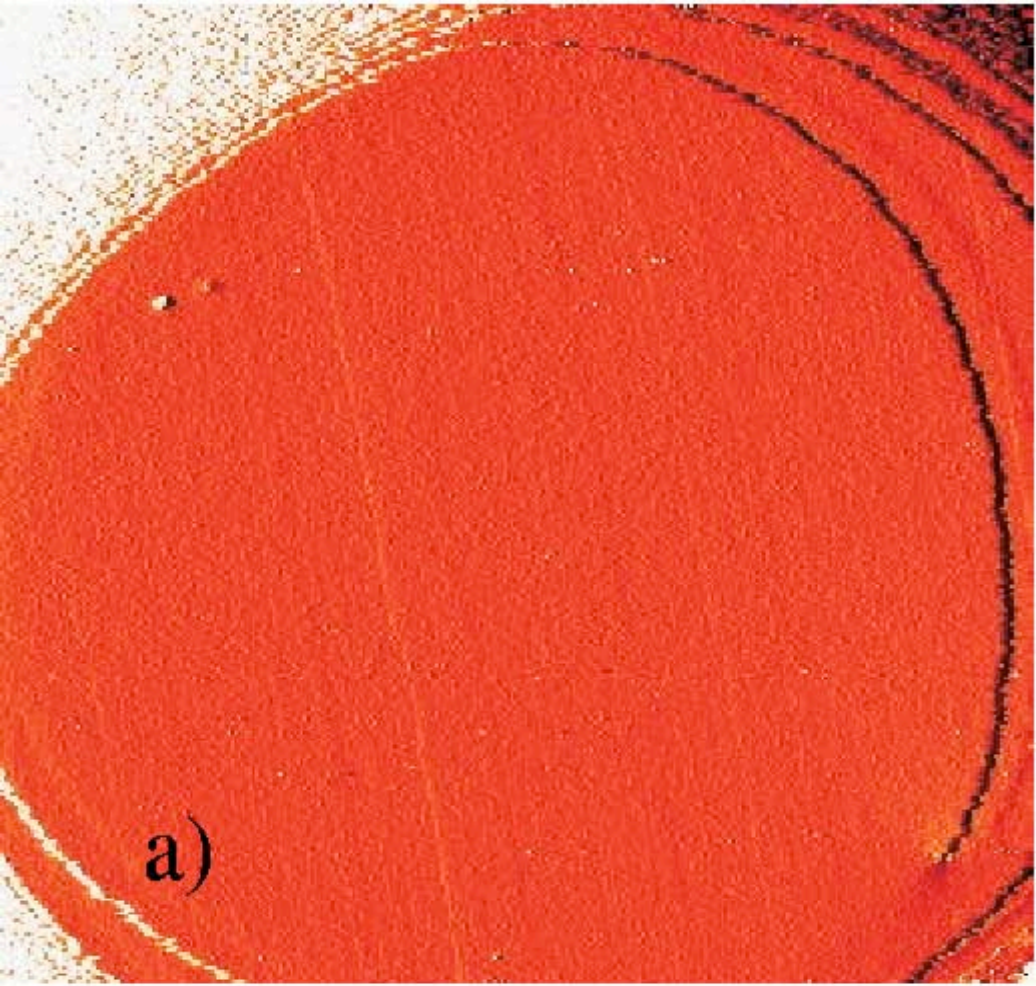

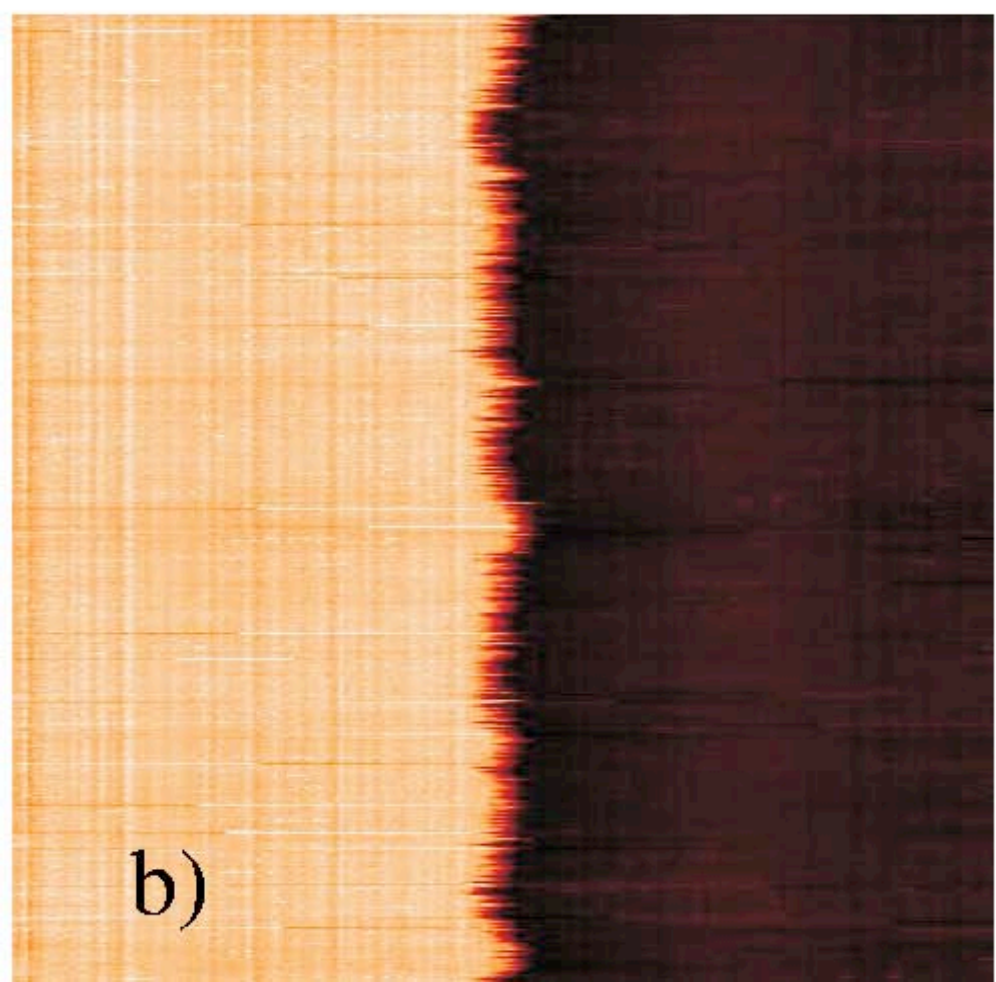

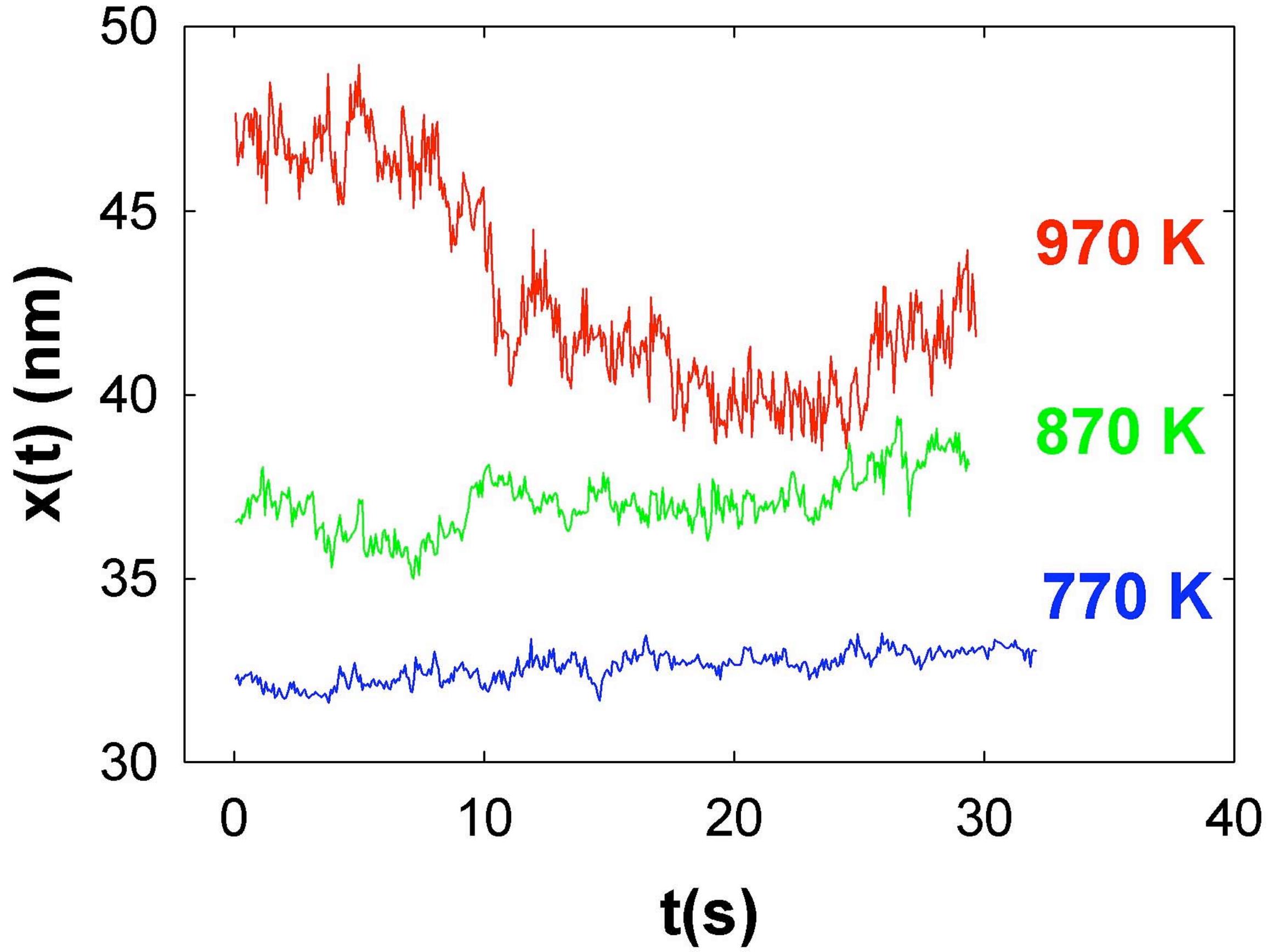

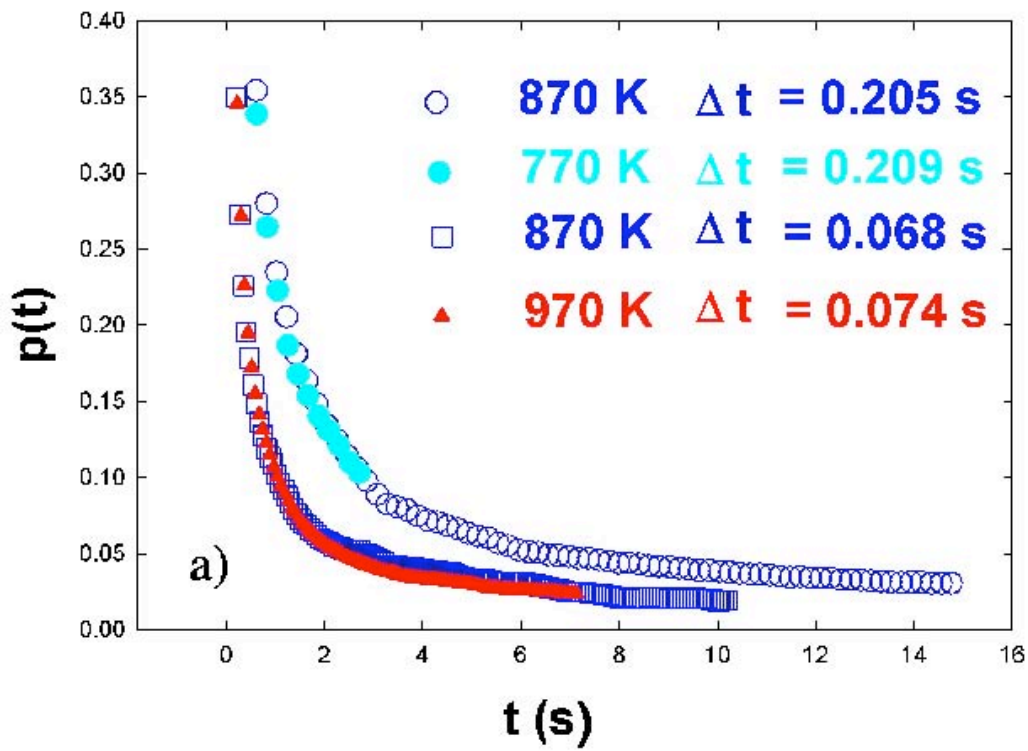

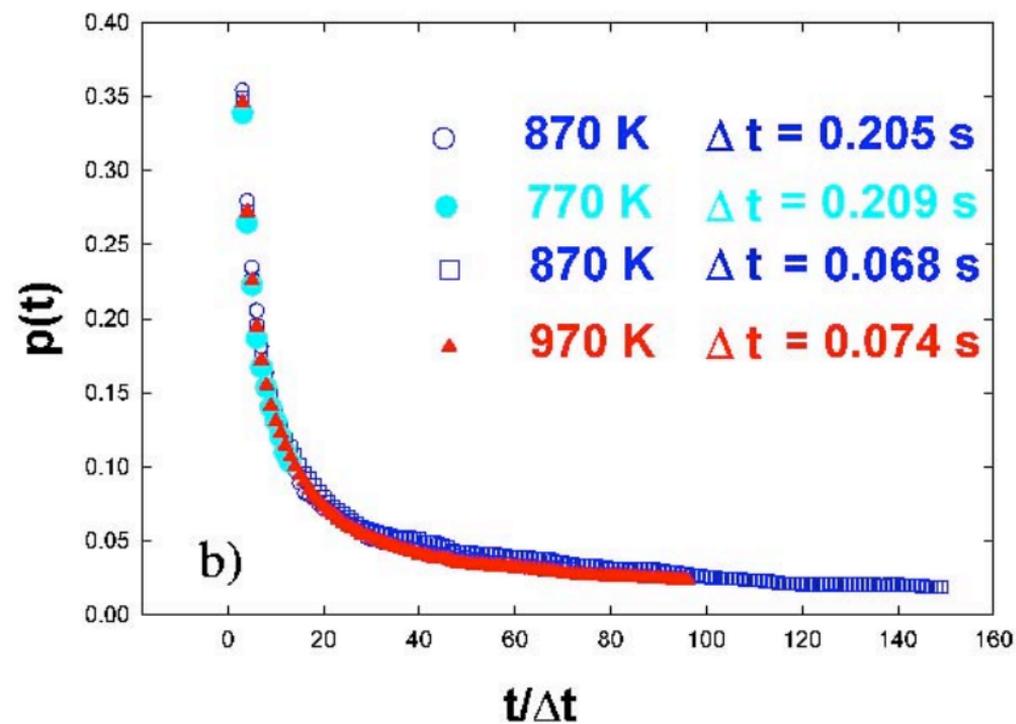

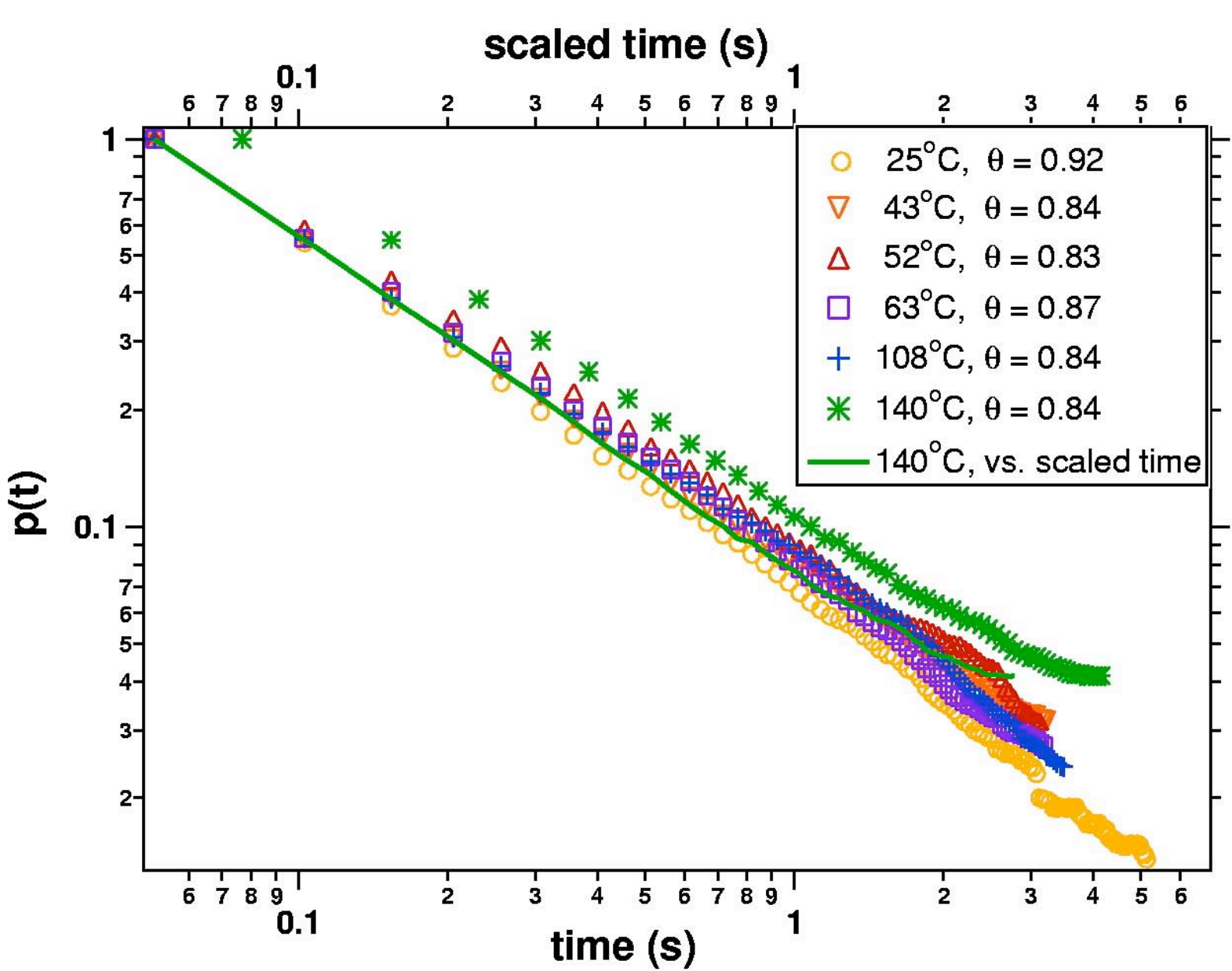

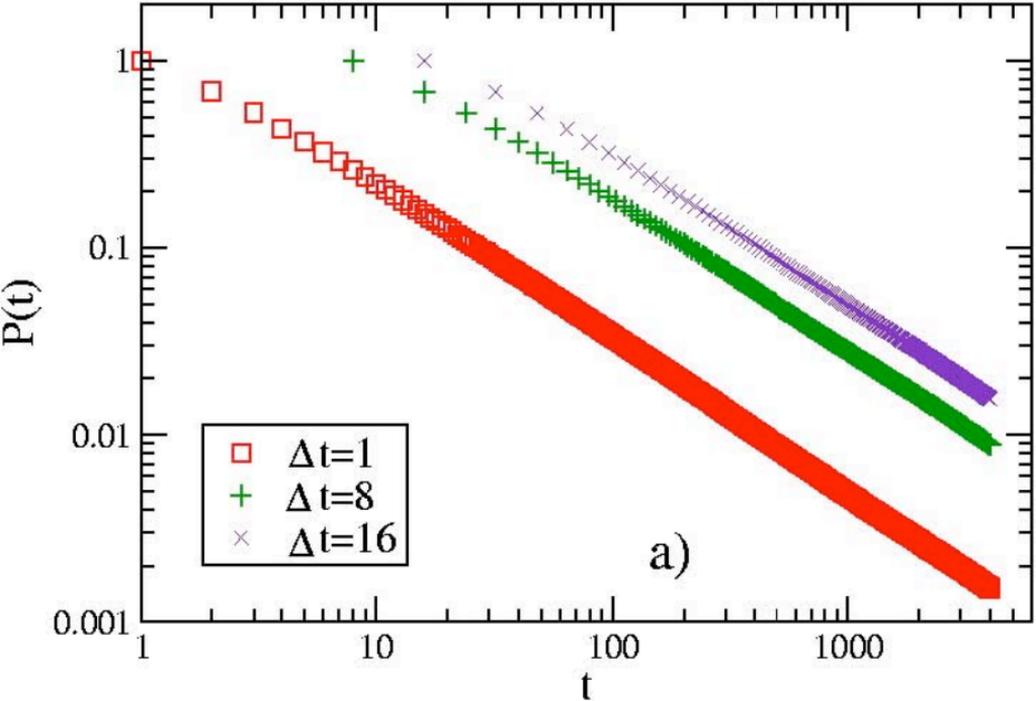

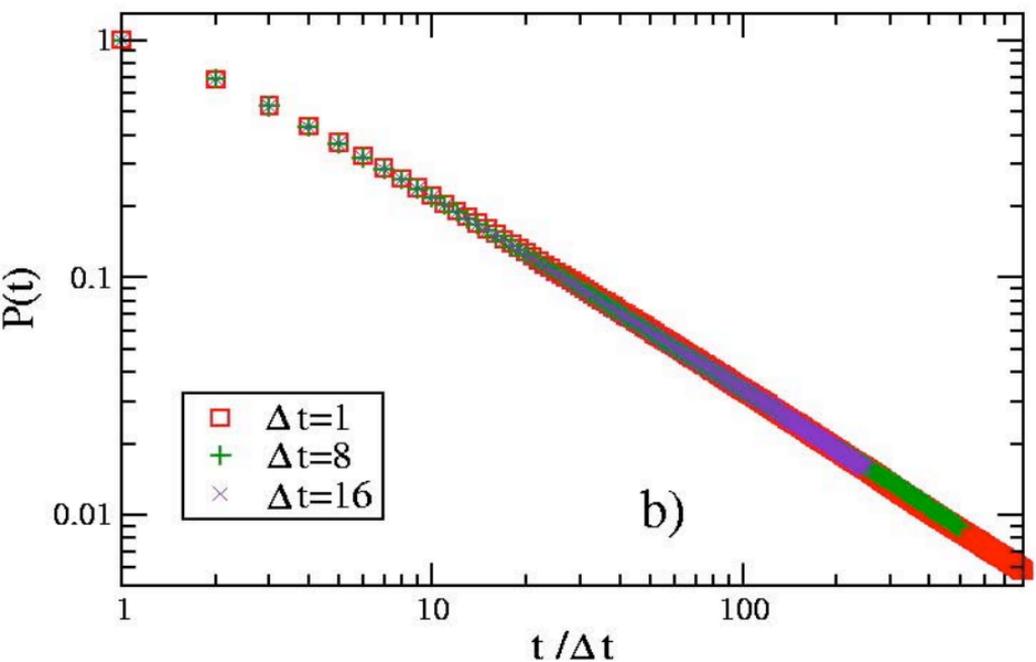

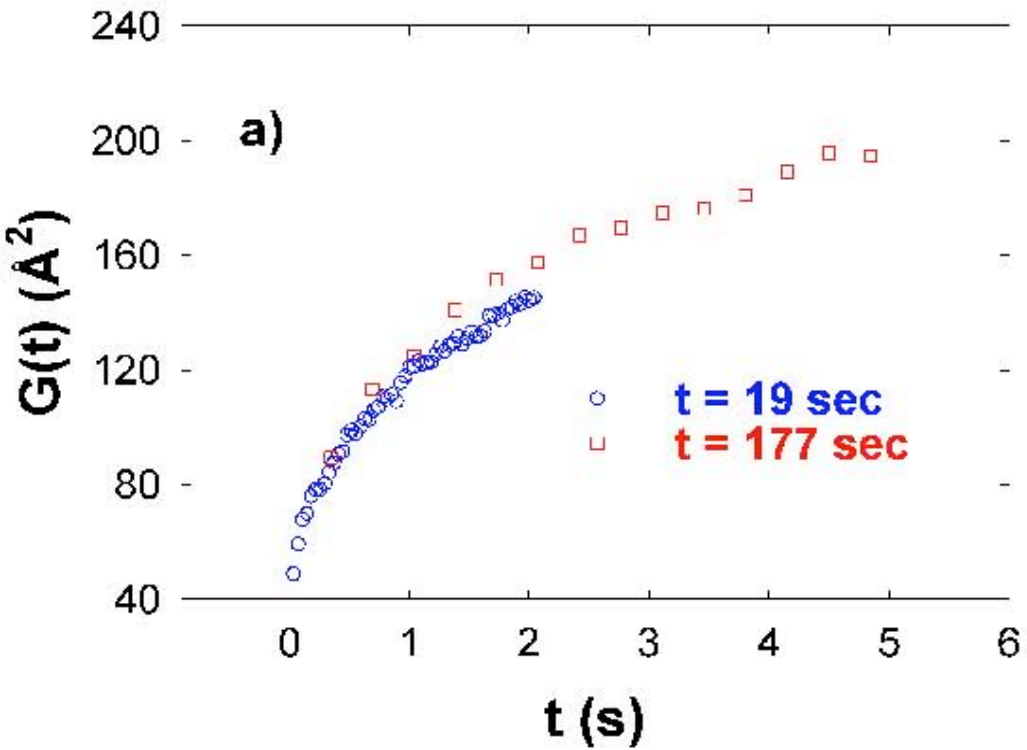

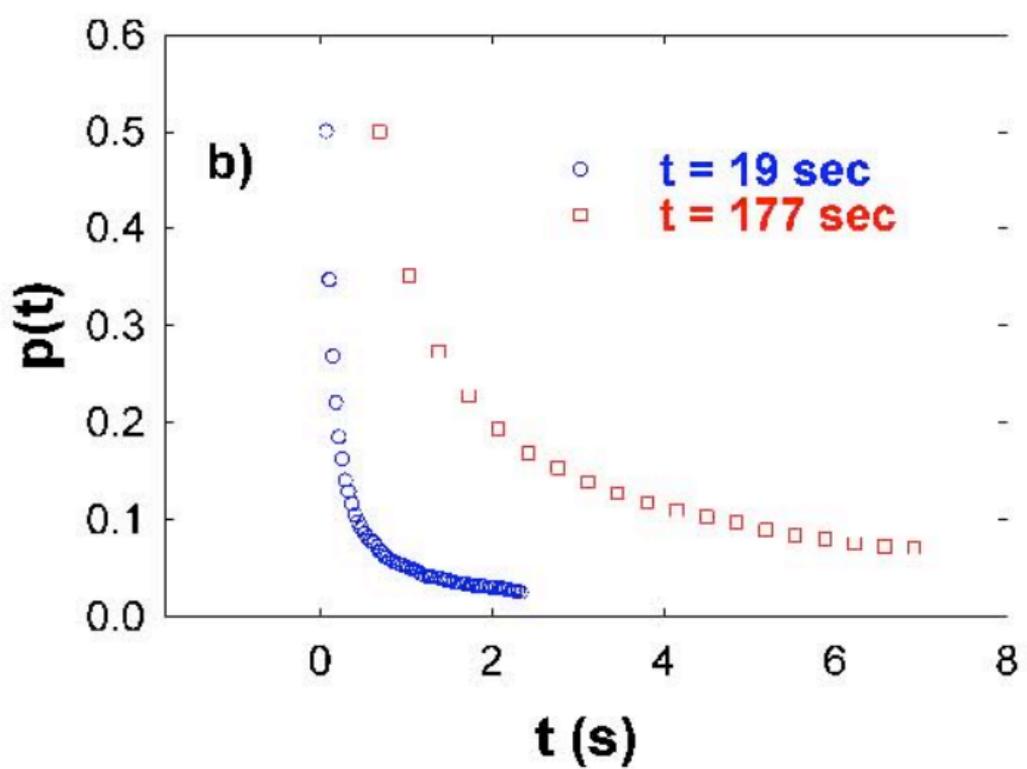

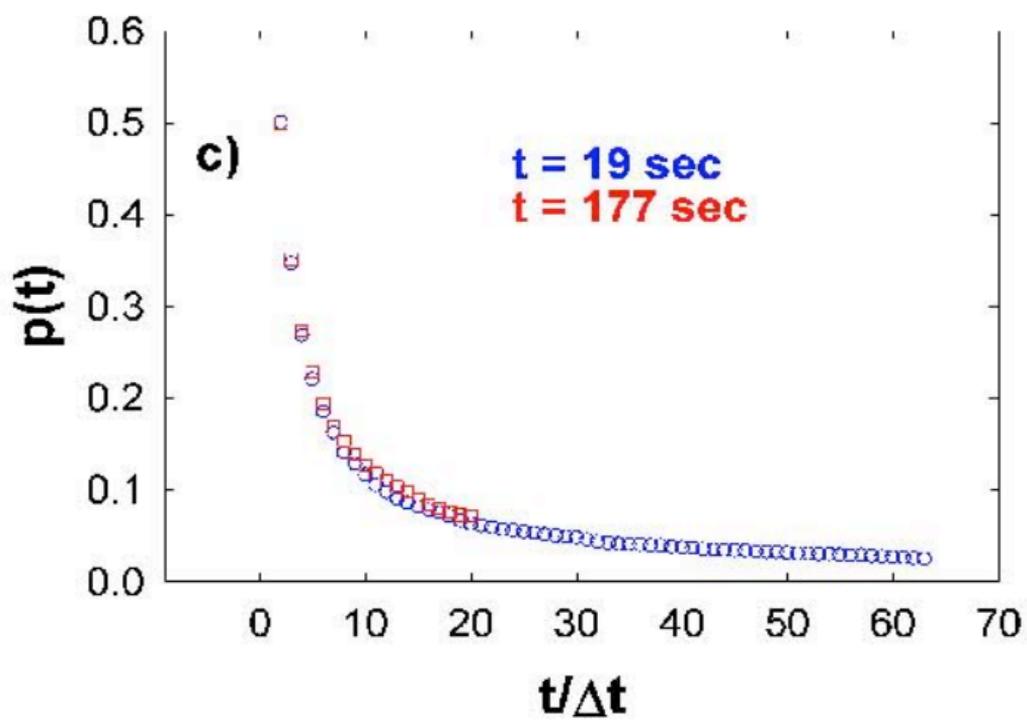

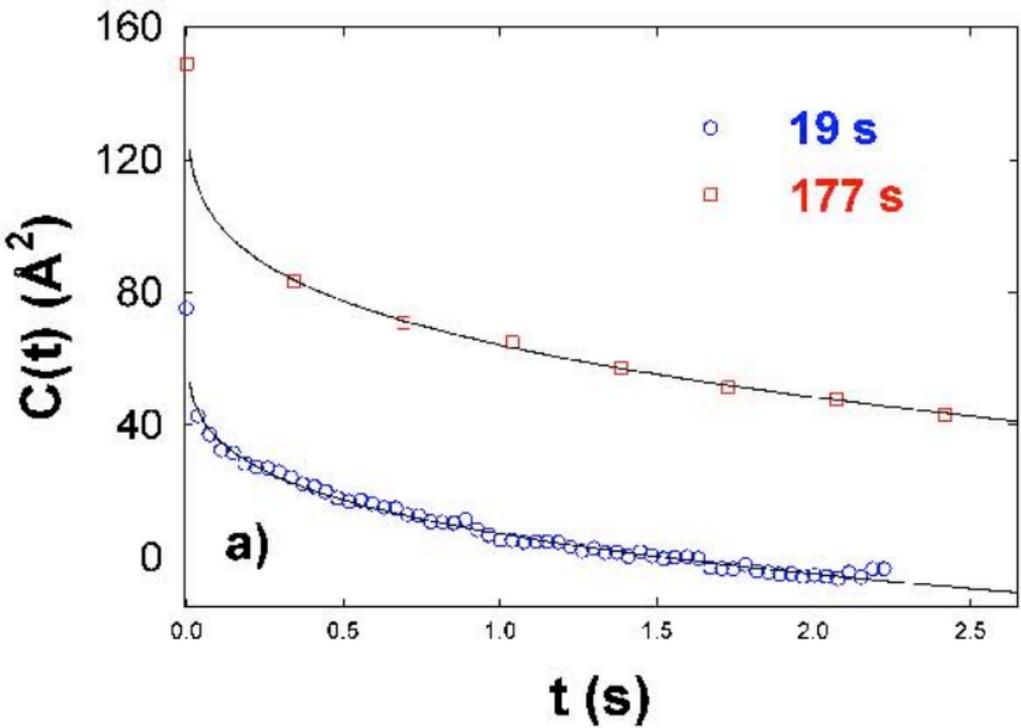

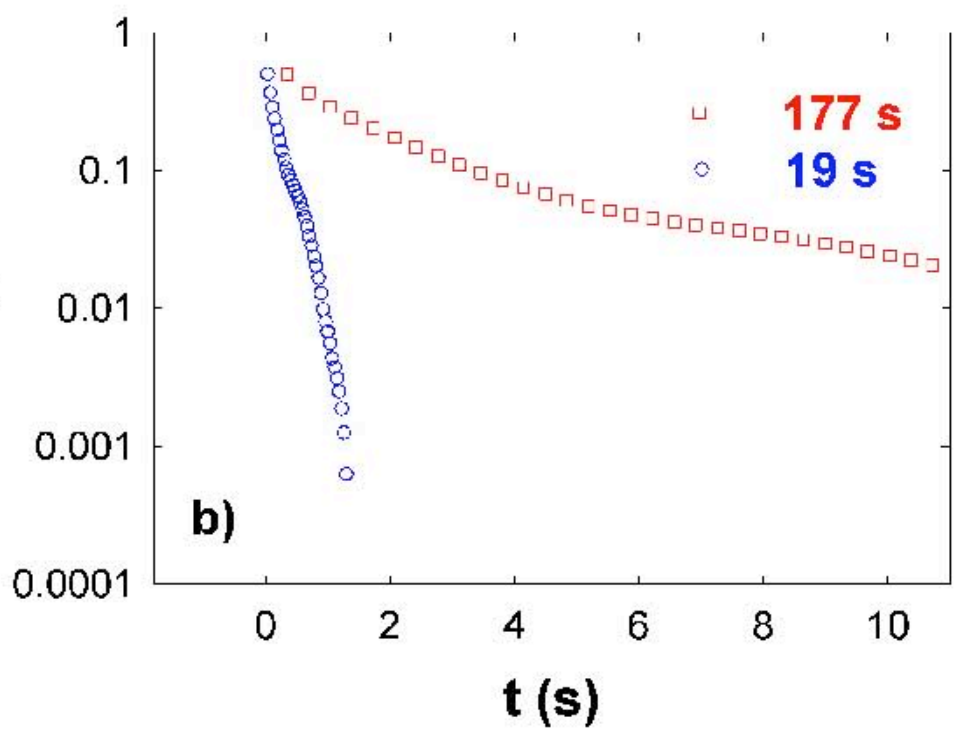

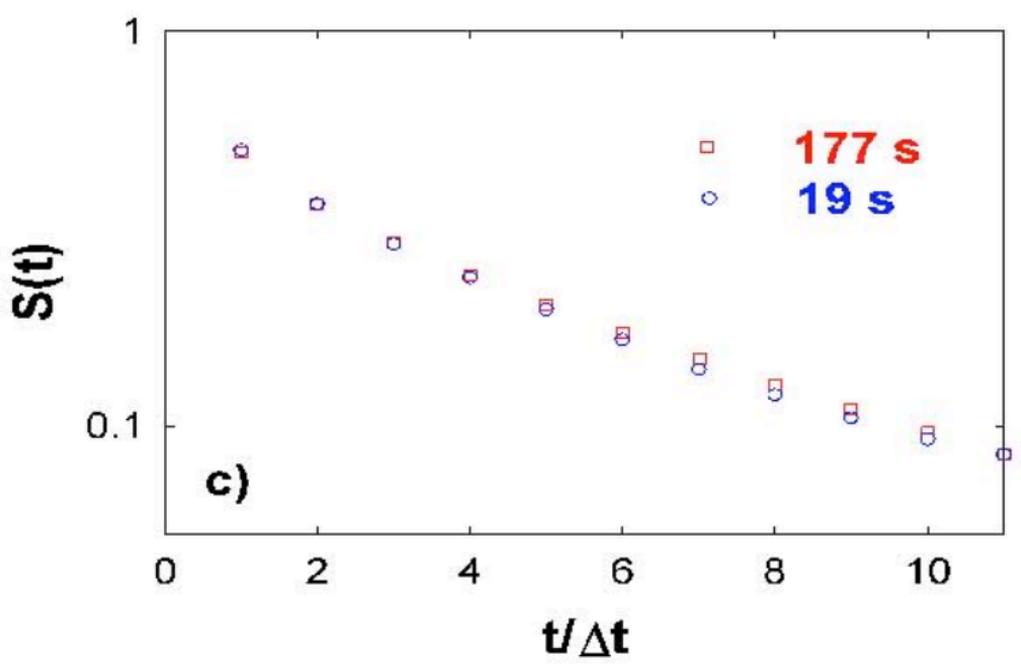

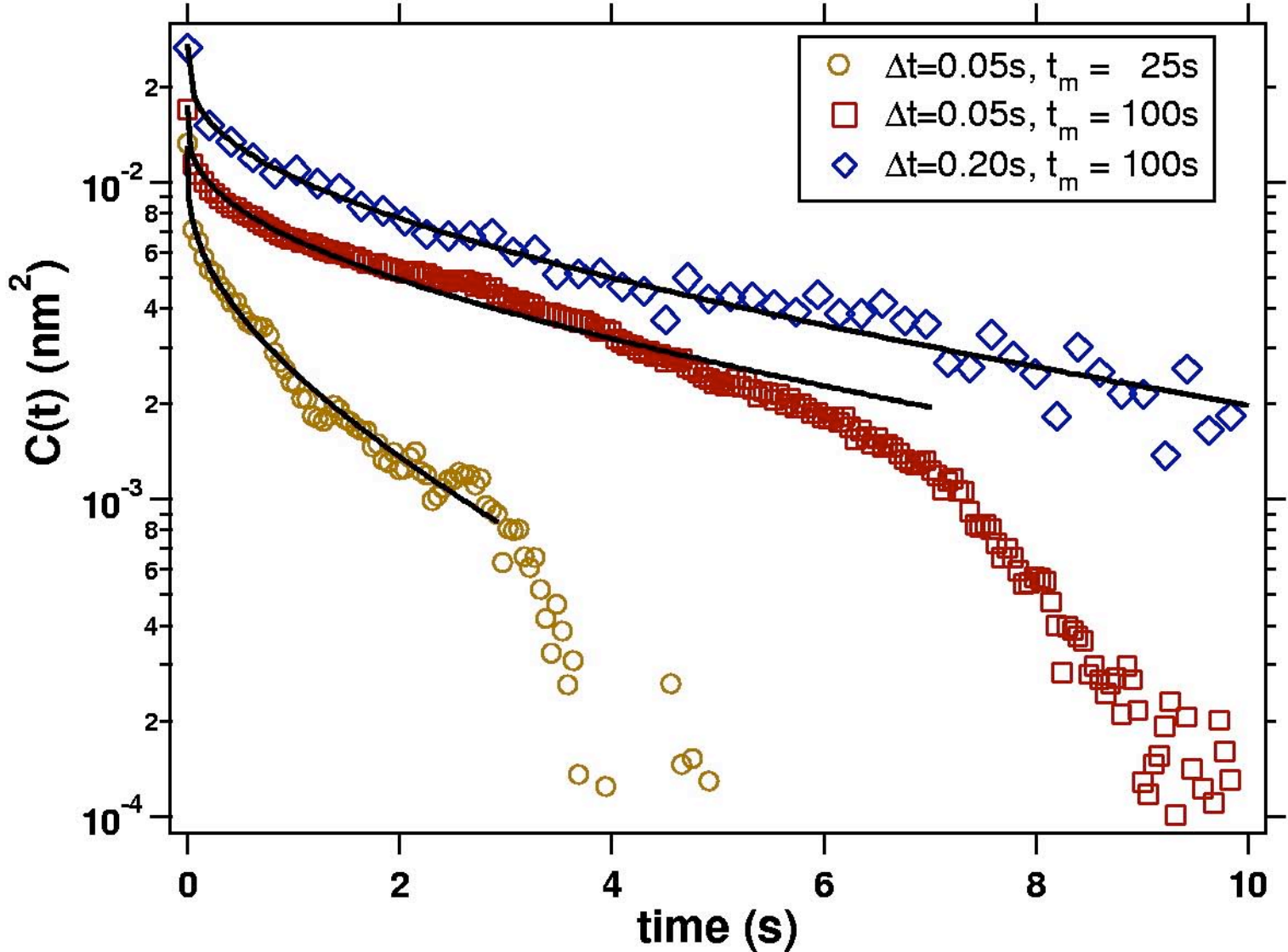

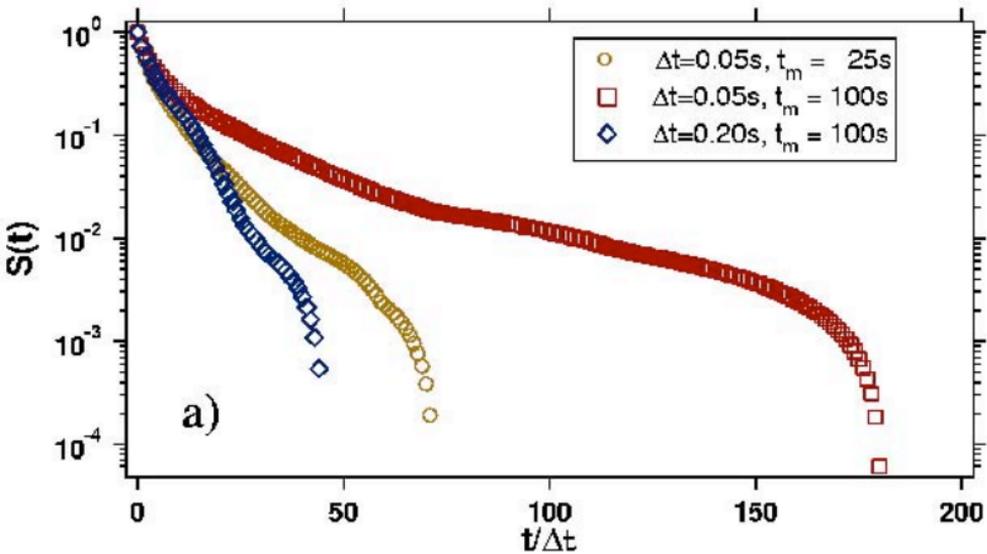

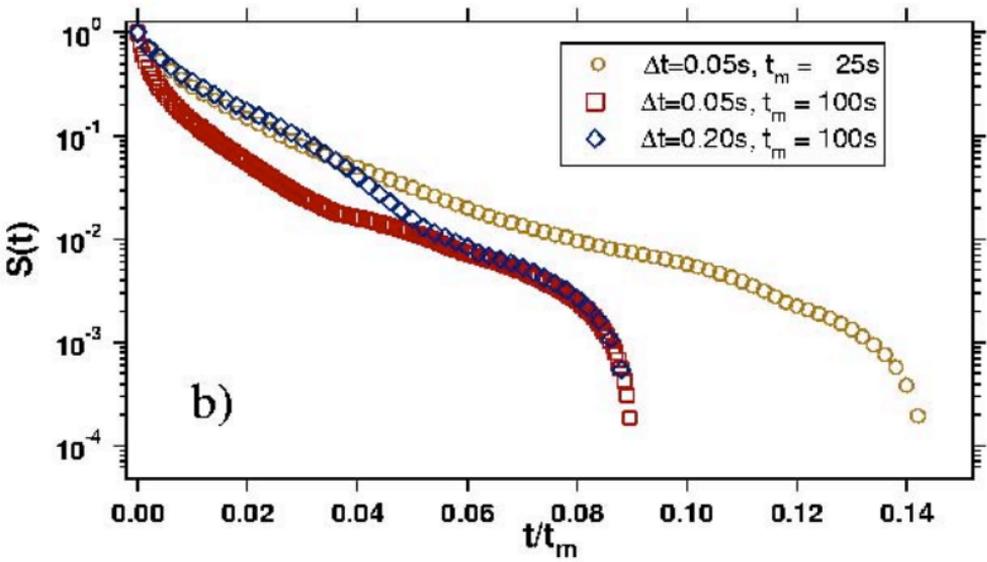

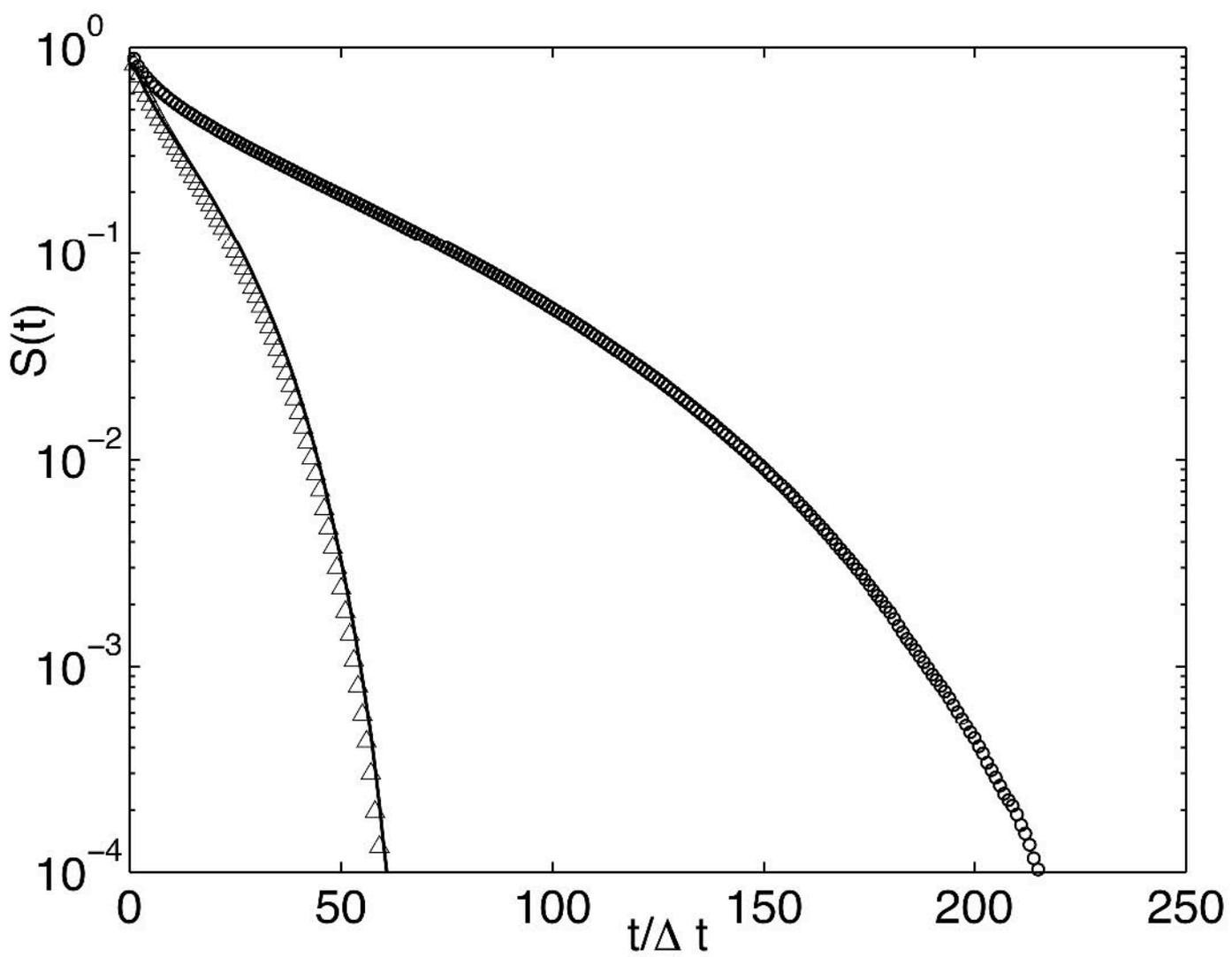